\documentstyle[prd,aps]{revtex}  

\def\be{\begin{equation}}
\def\ee{\end{equation}}

\begin{document}

\title{Gravitational field of a rotating gravitational dyon}

\author{Naresh Dadhich$^{a}$ and
Z. Ya. Turakulov$^{b}$\thanks{Permanent address:  Institute of Nuclear 
Physics,  Ulugbek,  Tashkent  702132, \,  Uzbekistan. }\\
{\sl Inter-University Centre for Astronomy \& Astrophysics,} \\
{\sl Post Bag 4, Ganeshkhind, Pune 411 007, India .}}

\maketitle

\begin{abstract}
  We have obtained the most general solution of the Einstein vacuum
  equation for the axially symmetric stationary metric in which both the 
  Hamilton-Jacobi equation for particle motion and the Klein - Gordon 
  equation are separable. It can be interpreted to describe the gravitational 
field of a rotating dyon, a particle endowed with both gravoelectric (mass) 
and gravomagnetic (NUT parameter) charge. Further, there also exists a 
duality relation between the two charges and the radial and the polar angle 
coordinates which keeps the solution invariant. The solution can however be 
transformed into the known Kerr - NUT solution indicating its uniqueness 
under the separability of equations of motion.
\end{abstract}

\vspace{0.5cm}


\vspace{0.5cm}

 It is well-known that the Kerr solution is the unique two parameter solution 
of the Einstein vacuum equation for asymptotically flat axially symmetric 
stationary spacetime admitting regular smooth horizon. The two parameters 
represent mass and angular momentum of a rotating black hole. This is by 
far the most interesting black hole solution of the Einstein equation with 
important applications in high energy astrophysics. The vacuum black hole 
could thus be specified by only two parameters. This fact is expressed in 
the relativistic literature by the famous statement that black hole has no 
hair.

 It is however worth enquiring, what happens if we drop asymptotic flatness? 
There do exist black hole spacetimes which are
not asymptotically flat. There are examples of the NUT and the Kerr - NUT
solutions [1,2], black holes sitting in uniform magnetic field [3] and in
cosmological models [4], and the black holes in the de Sitter universe. That 
is, dropping of the asymptotic flatness condition is not entirely 
unphysical and unknown. It is therefore a pertinent question to seek the most 
general solution without the assumption of asymptotic flatness. Among the 
examples cited above, the NUT and the Kerr - NUT are the only vacuum 
solutions. The NUT solution does indeed present a rather unfamiliar 
situation. It can be interpreted as describing the field of a gravomagnetic 
monopole charge [2,5]. It is however an interesting and novel solution. 

 The vacuum solutions are always interesting and more so the axially 
symmetric ones. A stationary vacuum spacetime is expected to be 
asymptotically flat representing a localized source. In spherical symmetry,
there cannot exist asymptotically non-flat vacuum solution [6]. On the other 
hand, axial symmetry however does not prohibit their existence. Do there 
exist such solutions other than the Kerr - NUT solution [2]? We shall in fact 
obtain the most general solution in a particular framework which is motivated 
by the solvability of basic physics equations. Our physical motivation is that 
the Hamilton - Jacobi (HJ) equation for particle motion and the 
Klein - Gordon (KG) equation must be separable in the variables of the   
metric which we seek as a solution of the Einstein equation. We would hence 
like to impose on 
the metric a priori the constraint of separability [7,8-10]. We then find the 
most 
general solution which can be transformed into the Kerr - NUT solution [2]. 
That means it is unique under the assumption of separability of the equations 
of motion. 

 The Kerr solution is the unique solution for a rotating particle which has 
gravoelectric charge (mass). What happens when the particle has gravomagnetic 
charge (NUT parameter), the rotating analogue of the NUT solution? Since the 
NUT solution is asymptotically non flat, its rotating analogue would also 
have to be so. That means we should seek the general solution of the vacuum 
Einstein equation by dropping the requirement of asymptotic flatness so as to 
let the particle have both gravo-(electric and magnetic) charge, a 
gravitational dyon. It 
turns out that the Kerr-NUT solution is the general solution for a dyon 
particle with both charge. 

 One of us had developed a method [7] of implementing the constraint of 
separability on the axially symmetric metric. Since the spacetime is 
stationary and axially symmetric, it would admit two familiar Killing vectors,
 and the metric could in general be written as,
\be
ds^2 = Bdt^2 - Ad\varphi^2 + 2 Cdt d\varphi - D \left(\frac{dr^2}{U^2} + 
\frac{d\lambda^2}{V^2}\right)
\ee
where $A, B, C, D$ are functions of $r$ and $\lambda$, and $U = U(r), 
V = V(\lambda)$. 

 Now we demand that HJ equation should be separable in $r$ and $\lambda$, and 
also, as in the case of Newman - Penrose fromalism, the coordinate surfaces 
corresponding to these two coordinates should contain null geodesics. This 
will lead to the requirement that the 
quantities, $AD/\rho^2, BD/\rho^2, CD/\rho^2, D$ are all functions of the type 
$X(r) + Y(\theta)$, where $\rho^2 = AB + C^2$ is the ($t-\varphi$) block 
determinant (for details, see [11]). Further, separability of KG equation will 
require $\rho^2 = UV$, which will also ensure the existence of horizon as a 
coordinate surface given by $U = 0$. With all this taken into account, We can 
then following [7] write the metric in the form [11],
\be
ds^2 = \Lambda(dt + \alpha d\varphi)^2 - (\Lambda)^{-1}\left[(U^2 - a^2V^2)
\left(\frac{dr^2}{U^2} + \frac{d\lambda^2}{V^2}\right) + U^2V^2 d\varphi^2\right] 
\ee
where 
\begin{eqnarray}
\Lambda &=& \frac{U^2 - a^2V^2}{F - a^2G}, \\
\alpha &=& a\frac{(F - U^2)V^2 - (G - V^2)U^2}{U^2 - a^2V^2}, \\
 F &=& r^2 + a^2, ~G = 1 - \lambda^2 
\end{eqnarray}
and $U = U(r), V = V(\lambda)$. Here $\lambda$ is an angle coordinate and $a$ 
is a 
constant having dimension of length. We shall refer this as the separable 
metric, and we shall find the most general solution of the Einstein vacuum 
equation for it.

 There are only two unspecified functions;  one of the radial coordinate, 
${\cal R}(r) = F - U^2$ and the other of the angular coordinate, 
$\Theta(\lambda) = G - V^2$. Now the most general vacuum solution is given by
\be
{\cal R}\equiv F - U^2 = 2Mr + (1 - k)a^2,  ~ \Theta\equiv G - V^2 = 
-2N\lambda + 1 - k  
\ee
where $M$ is a constant of dimension of length while $N$ and $k$ are
dimensionless constants. For the separable metric (2), the Ricci curvature 
has only five non zero components, and they readily yield the most general 
solution as given above [11]. The most remarkable feature of this general 
solution is that it remains invariant under the duality transformation, 
$M \leftrightarrow iaN, \, r \leftrightarrow ia\lambda$. Under this duality, 
$F \leftrightarrow a^2G, \, U^2 \leftrightarrow a^2V^2, \, {\cal R} 
\leftrightarrow \Theta$, clearly exhibiting the symmetry of the spacetime 
metric. 

 We define the coordinate $\theta$ by integrating for the angle variable 
$\lambda$ in the metric (2) and obtain
\be
\lambda = N + \sqrt{k + N^2} \cos\theta, ~V^2 = (k + N^2) \sin^2\theta.
\ee

In the Boyer - Lindquist coordinates, the metric (1) would read as,
\be
ds^2 =  \frac{U^2}{\rho^2}(dt - aG d\varphi)^2 - \frac{V^2}{\rho^2} 
(F d\varphi - a dt)^2 
- \frac{\rho^2}{U^2} dr^2 - \rho^2 d\theta^2 
\ee
where $\rho^2 = F - a^2G$. Note that when $\Theta = 0$, which would imply 
$N = 0, k = 1$, the 
solution reduces to the familiar Kerr solution with $M$ and $a$ having the 
usual meaning. On the other hand, when ${\cal R} = 0$ which implies $M = 0, 
k = 1$, it is again a non trivial vacuum solution which could be considered 
as dual to the the Kerr solution [12]. When $a = 0$, it reduces to the 
Schwarzschild solution with the deficit angle;i.e. $\sqrt{k + N^2}\varphi$ in 
place of $\varphi$. The angle deficit is the characteristic of the 
topological defect, cosmic string which causes a conical singularity [13]. 

 When both ${\cal R}$ and $\Theta$ vanish, it is the flat Minkowski 
spacetime. These are the two potential functions for the axially symmetric 
vacuum spacetimes, which are respectively anchored on gravoelectric ($M$) and 
gravomagnetic ($l=aN$) charges. The Kerr solution has gravoelectric charge 
while its dual has gravo-magnetic charge [2,5,12]. It is thus the duality 
of gravoelectric and gravomagnetic charges and fields, and the general 
solution is as shown above invariant under it.

 Further of the new dimensionless parameters $k, N$, it turns out that $k$ is 
removable while $N$ contributes to 
the Riemann curvature components only in presence of the parameter $a$, else 
it could only produce deficit angle. That means it has physical 
meaning only in conjection with the parameter $a$. This suggests that it 
should be possible to define new dimensionfull parameter, $l=aN$, which would 
turn out to be the NUT parameter. When we do that, the solution gets 
transformed to the Kerr - NUT solution [2].

 Write $k = (b^2 - l^2)/a^2, N = l/a, a^2 = d^2 + b^2 + l^2$, 
$\varphi = (a/b)\bar\varphi, t = \bar t + (d^2/b)\varphi$ and $\bar\alpha = 
(a\alpha +d^2)/b$, and then a straightforward calculation (on dropping 
overhead bars and finally replacing $b$ by $a$) takes the metric (2) to 
the eqn. (8) form,
\be
ds^2  =   \frac{U^2}{\rho^2}\left(dt - P d\varphi\right)^2 - \frac{\rho^2}{U^2}
dr^2 
- \frac{Q}{\rho^2}\left((\rho^2 + aP) d\varphi - a dt\right)^2 - \rho^2 
d\theta^2 
\ee
where
\be
F - U^2 = 2Mr + l^2, ~P - aQ = -2l \cos\theta, ~Q = \sin^2\theta.
\ee

 This is the Kerr - NUT solution [2] where $l$ is the NUT parameter 
representing the gravomagnetic monopole charge [2,5]. Thus the general 
solution we have found is actually the Kerr - NUT solution which is 
asymptotically non-flat. It is therefore the unique axially symmetric 
stationary vacuum solution for the separable metric in the form (2) or (8). It 
also contains the Kerr solution which is unique when the metric is 
asymptotically flat and admits regular event horizon. The present solution 
admits regular horizon but is not asymptotically flat. It is thus the non 
asymptotically flat generalization of the Kerr solution which is 
characterized by the separability of HJ and KG equations. \\

 We have thus obtained the most general vacuum solution for separable axially 
symmetric stationary spacetime. It describes the field of a 
rotating gravitational dyon, a particle having both electric
and magnetic gravitational charge. On further demand of asymptotic flatness, 
the magnetic charge gets switched off and  
it reduces to the familiar Kerr solution having only the electric charge, 
mass. It is interesting to note that asymptotic non flatness is therefore the 
characteristic property of the grovomagnetic charge, the NUT parameter $l$.   
Further note that the existence of regular horizon is implied by
separability of KG equation, while separability of HJ equation does not imply 
asymptotic flatness, and thereby leaving room open for a more general 
(than the Kerr) solution, which can also accommodate gravomagnetic charge. It 
turns out to be the known Kerr - NUT solution.

 We have reduced the parameters from four to three (and in 
particular $a, k, N$ to $b,l$, and then $b$ is written as $a$), now all of 
them are dimensionfull. That means all the three parameters, $a, k, N$ were 
not independent. However, the transformation to the Kerr - NUT form requires 
both $a$ and $b$ to be non-zero. When $a = 0$, the original metric reduces to 
the Schwarzschild solution with deficit angle while in the Kerr - NUT form it 
reduces to the NUT solution which would go to the Schwarzschild when $l = 0$. 

 In the Kerr - NUT form, when $P - aQ = 0$, it reduces to the Kerr solution 
while $F - U^2 = 0$ gives the dual to the Kerr solution. In the Kerr solution, 
the primary charge is gravoelectric denoted by $M$, while in the dual 
solution it is gravomagnetic denoted by the NUT parameter $l$. It is the 
rotation parameter $a$ which is responsible for gravomagnetic in the former 
and gravoelectric effects in the latter. These two solutions are therefore 
truly dual where their primary charge is interchanged [12]. 

 The electric charge could be easily included by replacing $2Mr$ in the 
general solution (6) by $2Mr - q^2$, where $q$ is the electric charge on the 
black hole. We have then verified that the scalar curvature $R$ vanishes and 
the stresses could be generated from the electromagnetic field of a rotating 
electric charge. When $a = 0$, it would reduce to the Reissner - 
Nordstr${\ddot o}$m solution. We have thus a four parameter family of 
electrovac black hole spacetime which is the non asymptotically flat or NUT 
generalization of the Kerr - Newman black hole. This is the most general
electrovac black hole solution [11]. The Vaidya generalization of a black hole 
solution, in which there is a 
radial null radiation flux, could be accomplished by transforming the black 
hole metric into the Eddington coordinates, and then taking $M$ as a function 
of the retarded time. The same procedure could be implemented for the metric 
(9) as well to get its radiating generalization [11]. The stresses conform to 
the Vaidya null fluid asymptotically.

 We have shown that, like the Kerr solution, the Kerr-NUT solution 
is also unique for the separability of HJ and KG equations. In other words, we
 have identified the conditions for uniqueness of the Kerr - NUT solution. 
The NUT family is the gravomagnetic analogue of the familiar gravoelectric 
Kerr family, and the two are dual of each-other. The uniqueness of the Kerr - 
NUT family is therefore very 
pertinent and important because it describes the field of a gravitational 
dyon. The most remarkable feature of the general solution 
is the duality symmetry of the spacetime metric (2-5) which exhibits the 
duality relation between the mass and the NUT parameters. This is a very 
interesting and important feature which is being exhibited for the first 
time. The rotation parameter, $a$, plays the key facilitative role in the 
duality transformation. In its absence, Schwarzschild and NUT solutions 
exhibit no such duality. That means the NUT 
parameter by itself cannot be dual to gravoelectric charge, mass. It acquires 
that character only when it is rotating. It would be interesting to study 
physical properties of the dual Kerr spacetime, and in particular quantum 
fields and the effects they might give rise to.

 As the Kerr solution is unique for a rotating black hole, the Kerr - NUT 
solution is thus unique for a rotating dyon black hole endowed with 
both electric and magnetic gravitational charge.
 
{Acknowledgment:} We wish to thank Parampreet Singh for his help in verifying 
the solution at various stages of its evolution. ZYT  wishes to thank ICTP, 
Trieste for a travel grant under the BIPTUN program and IUCAA for warm 
hospitality. This result was first presented by one of us (ND)in the 
Workshop on {\it Super String Theory}, April, 23 - May, 12, 2001, 
Quaid-i-Azam University, Islamabad.

\end{document}